# An Efficient Fuzzy Clustering-Based Approach for Intrusion Detection

Huu Hoa Nguyen, Nouria Harbi and Jérôme Darmont

Université de Lyon (ERIC Lyon 2) - France
nhhoa@eric.univ-lyon2.fr, {nouria.harbi, jerome.darmont}@univ-lyon2.fr

**Abstract.** The need to increase accuracy in detecting sophisticated cyber attacks poses a great challenge not only to the research community but also to corporations. So far, many approaches have been proposed to cope with this threat. Among them, data mining has brought on remarkable contributions to the intrusion detection problem. However, the generalization ability of data mining-based methods remains limited, and hence detecting sophisticated attacks remains a tough task. In this thread, we present a novel method based on both clustering and classification for developing an efficient intrusion detection system (IDS). The key idea is to take useful information exploited from fuzzy clustering into account for the process of building an IDS. To this aim, we first present cornerstones to construct additional cluster features for a training set. Then, we come up with an algorithm to generate an IDS based on such cluster features and the original input features. Finally, we experimentally prove that our method outperforms several well-known methods.

## 1 Introduction

In recent years, with the dramatically increasing use of network-based services and the vast spectrum of information technology security breaches, more and more organizational information systems are subject to attack by intruders. Among many approaches proposed in the literature to deal with this threat, data mining brings on a noticeable success to the development of high performance intrusion detection systems (IDSs). The preeminence of such an approach lies in its good generalization abilities to correctly classify (or detect) both known and unknown attacks. However, as an inherent essence, the effectiveness of data mining-based IDSs depends heavily upon the quality of IDS datasets. In practice, IDS datasets are often extracted from raw traces in a chaotic system environment, and hence could hold implicit deficiencies, e.g., the existence of noise in class labels due to mistakes in measurement, and the lack of base features. Moreover, due to the sophisticated characteristics of attacks and the diversification of normal events, different data regions could behave differently, i.e., true class labels could seriously be interlaced.

Such factors pose a great difficulty for inducers to identify appropriate decision boundaries from the input space of IDS datasets. In other words, when the input space is not robust enough to discriminate class labels, making further treatments from alternative knowledge sources as new supplemental features is highly desirable. To this aim, one common approach is to transform the input space into a higher dimensional space from which data are more separable. New additional features can be found by either manual ways based on prior knowledge or automatic analysis methods (e.g., principle component analysis). However, in a high dimensional input space, finding new relevant features is a tough task that often requires human analyses, but derived features are sometimes not as good as expected. As a result, in practice, one often applies standard dimensional-transformation methods (e.g., polynomial, radial basic function) to application domains where class discrimination is ambiguous and additional features are hard to be identified. Yet, such methods are greatly affected by input parameters and data distribution, thus not always outputting a high performance classifier. In this vision, it is desirable to find additional features in a less complex way so that general-purpose algorithms such as Decision Trees (DT) or Support Vector Machines (SVM) can learn the data more efficiently.

Such a context motivates us to propose a novel approach that treats fuzzy cluster information as additional features. These features are selectively incorporated into the input space for building an efficient IDS. we experimentally show that our solution approach is considerably superior to several well-known methods.

The remainder of this paper is organized as follows. Section 2 presents the problem formulation of our approach, whereas section 3 describes our solution for generating an IDS. Section 4 shows the experimental results we achieved. Section 5 finally gives a conclusion of the method we propose.

## 2 Problem formulation

Clustering aims to organize data into groups (clusters) according to their similarities measured by some concepts. Unlike crisp clustering that crisply assigns each data point to a separate cluster, fuzzy clustering allows each data point to belong to various clusters with different membership degrees (or weights). Fuzzy clusters are expressed by their centers (or centroids) that are simultaneously found in the partitioning process of a fuzzy clustering algorithm. The number of clusters ($k$) is often inputted as a parameter to a fuzzy clustering algorithm. The $n \times k$ membership matrix $W=\{w_{ij} \in [0,1]\}$ of $n$ data points is found in the fuzzy clustering process. For example, Figure 1 describes the instance space of a training set partitioned into four fuzzy clusters, where membership weights that data point $x_1$ belongs to clusters '1', '2', '3', and '4' are 0.3, 0.14, 0.16, and 0.4, respectively.

Let us first denote $S=\{X,Y\}$ the original training set of $n$ data points $X=\{x_1,\ldots,x_n\}$, where each point $x_i$ is an $m$-dimensional vector $(x_{i1},\ldots,x_{im})$ and assigned to a label $y_i \in Y$ belonging one of the $c$ classes $\Omega=\{\omega_1, \ldots,\omega_c\}$. Let $B=\{b_i|\ b_i=\max(w_{ij}), j=1\ldots k\}$ hold the maximum membership weight of each point $x_i$, and $Z=\{z_i|\ z_i=\mathrm{argmax}_j(w_{ij}), j=1\ldots k\}$ contains the cluster (symbolic) number assigned to each point $x_i$.

For conciseness in describing the approach, we term two column matrices $Z$ and $B$ as two "basic cluster features". In addition, we name the $j^{th}$ column of the membership matrix ($W$) as $P_j$, and term the columns $P_1, \ldots, P_k$ as "extended cluster features". We also term the training set added with cluster features $\{X, Z, B, P_1, \ldots, P_k, Y\}$ as a "manipulated training set". These notations and terminologies are depicted in Figure1.

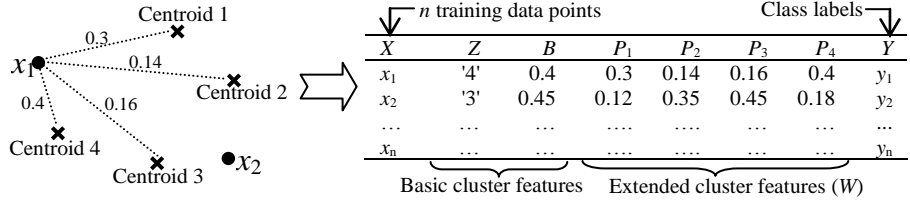

**Fig. 1.** A manipulated training set, resulting from adding cluster features into the input space.

The problem formulation follows: "Given a training set $S=\{X,Y\}$ and an inducer $I$, the goal is to find a high performance classifier induced by $I$ over the $m$ initial features of $S$ and the supplemental cluster features $\{Z, B, P_1, P_2, \ldots, P_k\}$ resulting from a parameterized-by-$k$ fuzzy clustering based on $X$".

Undoubtedly, fuzzy clustering has a great potential in expressing the latently natural relationships between data points. Here, a question is whether information about fuzzy clusters benefits certain inducing types. Basically, there exist some types of inducers to which fuzzy cluster features are helpful. For example, in the SVM context, the decision boundary often falls into a low density region, but the true boundary might not pass through this region, thus resulting in a poor classifier. However, when supplemented with relevant cluster features, data points in high dimensional spaces can become more uniform and discriminatory, hence avoiding an improper separation across this region. In fact, the crucial factor to the success of SVM lies in a kernel trick that maps the initial input space to a much higher dimensional feature space, where the transformed data are expected to be more separable from a linear hyper-plane function. In order words, while other inducers somewhat find dimensionality a curse, blessing of dimensionality can enable SVM to be more effective. Under such a sense, incorporating relevant cluster features into the input space discernibly benefits SVM inducers.

Another consideration relates to the univariate Decision Tree (DT) setting. Due to its greedy characteristic, the DT inducer examines only one ahead partitioning step for growing child trees, rather than considering deeper partitioning steps that can achieve a better tree. This characteristic can lead to an improper tree-growing termination (e.g., the XOR problem), and thus generate a poor classifier. In this vision, cluster features help the DT inducer to determine splits more properly for tree growing.

## 3 Fuzzy Cluster Feature-based Classification

### 3.1 Cluster Feature Generation and Selection

Basically, cluster features can be generated by any fuzzy clustering algorithm. However, for concreteness, we express cluster features with the fuzzy $c$-means clustering [6], which typically solves the minimization problem to the objective function of Formula 1. In a common form, the objective function (Formula 1) reaches to a minimum over $W$ (membership matrix) and $V$ (centroids), by Formulas 2 and 3.

$$f_{obj}(X,W,V) = \sum_{j=1}^{k}\sum_{i=1}^{n} w_{ij}^{\alpha} d(x_i,v_j)^2 \text{, subject to the constrain } \sum_{j=1}^{k} w_{ij} = 1 \quad (1)$$

$$v_j = \left(\sum_{i=1}^{n}(w_{ij})^{\alpha} x_i\right) \bigg/ \left(\sum_{i=1}^{n}(w_{ij})^{\alpha}\right) \quad (2)$$

$$w_{ij} = \left(\frac{1}{d(x_i,v_j)^2}\right)^{\frac{1}{\alpha-1}} \bigg/ \sum_{q=1}^{k}\left(\frac{1}{d(x_i,v_q)^2}\right)^{\frac{1}{\alpha-1}} \quad (3)$$

where $\alpha$ is a fuzzy constant and $d(x_i,v_j)$ is the distance from $x_i (\in X)$ to $v_j (\in V)$

Fuzzy $c$-means clustering tries to find the best fit for a fixed value of $k$, the number of clusters. However, as an essential problem of clustering, determining an appropriate parameter $k$ is a tough task. The most common way to find the reasonable number of clusters is to run the clustering with various values of $k \in \{2,\ldots, k_{max}\}$ and then use a validity measure (e.g., partition coefficient) to evaluate cluster fitness.

In our approach, however, we need data to be grouped in a way that reveals helpful information for inducers, not for clustering itself, even though the number of clusters might be wrong. In other words, using validity measures to determine the best number of clusters is not reliable enough to derive good cluster features for classifiers. In such a vision, instead of endeavoring to find the best $k$ with validity measures, we use the *over-production* method to generate several candidate classifiers for different values of $k$ and then evaluate their performance to determine the best one. Evaluating the performance of candidate classifiers can be based either on a validation set or Cross Validation (CV) method [9]. Thus, a proper value of $k$ is simultaneously found in the process of finding a maximum performance classifier from candidate classifiers.

In addition, the use of cluster features should be examined individually for a concrete inducing type. Intuitively, two basic cluster features ($Z$, $B$) are benefic enough for DT inducer, instead of including $k$ extended cluster features ($P_1,\ldots,P_k$). By contrast, in the SVM context, it is applicable to employ either only the basic cluster features ($Z$, $B$) or all the cluster features ($Z$, $B$, $P_1,\ldots,P_k$) for building a classifier. Another solution that can be applied for any inducing type is to employ feature selection techniques (e.g., filter, wrapper) to pick out high merit features from both $m$ initial input features and all ($k+2$) cluster features. The objective is to apply feature selection techniques on ($m+k+2$) features to bring about a smaller but more qualitative feature subset than those only on $m$ initial features. Here, note is that feature selection is simultaneously carried out in the process of building candidate classifiers. In a nutshell, formally, there are three possibilities to incorporate cluster features into the initial features ($A_1, \ldots, A_m$), i.e., ($A_1, \ldots, A_m, Z, B$), ($A_1, \ldots, A_m, Z, B, P_1, \ldots, P_k$), or *Feature Selection*($A_1, \ldots, A_m, Z, B, P_1, \ldots, P_k$).

### 3.2 Algorithm for generating a fuzzy cluster feature-based classifier

Our algorithm for generating a classifier from both initial and cluster features, called *CFC*, is depicted from Figure 2. Related notations are indicated in Table 1.

**Table 1.** Notations used in Figure 2.

| Notation | Description |
|---|---|
| $C_k$ | A candidate classifier resulting from a clustering with $k$ fuzzy clusters. |
| $C_{k*}$ | The best classifier among $|K|$ candidate classifiers. |
| $V^k$ | A $k \times m$ matrix of $k$ centroids obtained from clustering $X$ into $k$ clusters. |
| $V^{k*}$ | A $k* \times m$ matrix of $k*$ centroids, corresponding to $C_{k*}$. |
| $W^k$ | An $n \times k$ membership matrix of $n$ data points $x_i \in X$, corresponding to $V^k$. |
| $B^k$ | A column matrix containing the maximum membership weight of each $x_i \in X$. |
| $Z^k$ | A column matrix representing the cluster (symbolic) number of each $x_i \in X$. |
| $\Theta$ | A horizontal concatenation operator between two matrices. |

*Training phase*
**Input:** $S=\{X, Y\}$: The original training set
      $I$: a base inducer
      $K$: a predefined integer set representing possible number of clusters
      $\xi$: a feature selection technique that returns a specific feature subset
      $T$: a type to employ features for building classifiers
**Output:** $C_{k*}$, $V^{k*}$

1: $X' \leftarrow \text{Normalize}(X)$     //Normalize continuous features
2: *For* each $k \in K$ do
3:    $\{W^k, V^k\} \leftarrow \text{FuzzyClustering}(X', k)$
4:    $B^k \leftarrow \{b_i \mid b_i = \max(w_{ij}), i=1...n, j=1...k\}$
5:    $Z^k \leftarrow \{z_i \mid z_i = \arg\max_j(w_{ij}), i=1...n, j=1...k\}$
6:    *Case*     //$D$ is a manipulated training set
7:       $T=1$: $D \leftarrow (X \Theta Z^k \Theta B^k)$   //Initial features & basic cluster features
8:       $T=2$: $D \leftarrow (X \Theta Z^k \Theta B^k \Theta W^k)$   //Initial features & all cluster features
9:       $T=3$:
            $F \leftarrow \xi(X \Theta Z^k \Theta B^k \Theta W^k, Y)$   //Apply a feature selection
            $D \leftarrow (X \Theta Z^k \Theta B^k \Theta W^k)[F]$   //Project data by the derived subset
10:   *End Case*
11:   $C_k \leftarrow I(D, Y)$   //Build a classifier, using the manipulated training set $D$ & inducer $I$
12:   $\text{Performance}(C_k) \leftarrow \{\text{Average performance of } q\text{-fold CV based on } (D,Y) \text{ and } I\}$
13: *End For*
14: $C_{k*} \leftarrow \arg\max_{C_k} \text{Performance}(C_k), k \in K$   //Determine one best classifier
15:   Return $C_{k*}$, $V^{k*}$

*Operation phase*
16: For an unlabeled testing instance $x$:
17: $x' \leftarrow \text{Normalize}(x)$     //Normalize continuous features
18: Compute membership weights $(w_j \mid j=1...k*)$ that $x'$ belongs to $v_j \in V^{k*}$ (Formula 3)
19: $b \leftarrow \max(w_j \mid j=1...k*)$
20: $z \leftarrow \arg\max_j(w_j \mid j=1...k*)$
21: Label $x$, by taking cluster features $\{z, b, w_j\}$ into account, using $C_{k*}$

**Fig. 2.** Algorithm *CFC*.

The key idea is that, for each clustering with different number of clusters ($k \in K$), the algorithm builds and valuates a candidate classifier from the training set manipulated with a given feature selection type, by $q$-Fold Cross Validation [9]. The resulting classifier is the one exhibiting maximum performance.

In the training phase, the algorithm first normalizes continuous features (e.g., by a variance-based spread measure) to avoid the dispersion in different ranges (Line 1). Here, it is noticed that the normalized data ($X'$) is merely for clustering purpose, whereas classifiers are built by using the original data ($X$). In addition, instead of executing clustering with parameter $k$ ranging from 2 to a given $kmax$ value, the algorithm uses a predefined set $K=\{k\}$ to mainly focus on important values of $k$, which can be recognized by experiment or prior knowledge (Line 2). As mentioned in Section 3.1, there are three cases to incorporate cluster features into the initial features. Hence, for general purpose, the algorithm introduces an input parameter $T$ for specifying the way to employ features for building classifiers (Lines 6-10). Subsequently, the algorithm builds and evaluates one candidate classifier for each clustering (Lines 11, 12). Here, note is that evaluating candidate classifiers is based on the averaged performance of $q$-fold stratified cross validation from the manipulated training set. Finally, the algorithm determines one best classifier from $|K|$ candidate classifiers, together with a corresponding centroid set (Lines 14, 15).

In the operation phase, for an unlabeled testing instance $x$, the algorithm first normalizes $x$ in the same way as those applied to the training set. Then, cluster features of $x$ are calculated based on the centroid set $V^{k^*}$ (Lines 18-20). Finally, the corresponding features are input to classifier $C_{k^*}$ for final prediction (Line 21).

## 4 Experiments

### 4.1 Dataset

Our experiments are conducted on the intrusion detection dataset KDD99 [3]. This dataset was derived from the DARPA dataset, a format of TCPdump files captured from the simulation of normal and attack activities in the network environment of an air-force base, created by MIT's Lincoln Laboratory. The KDD99 dataset comprises 494,021 training instances and 311,029 testing instances. Due to data volume, the research community mostly uses small subsets of the dataset for evaluating IDS methods. Each instance in the dataset represents a network connection, i.e., a sequence of network packets starting and ending at some well defined times, between which data flows to and from a source IP address to a target IP address under some well defined protocol. Such a connection instance is described by a 41-dimensional feature vector and labeled with respect to five classes: *Normal*, *Probe*, *DoS* (denial of service), *R2L* (remote to local), and *U2R* (user to root).

To facilitate experiments without losing generality, we only use a smaller set of the KDD99 dataset for the purpose of evaluating and comparing our method to others. In particular, the training and testing sets used in our experiments are made up of 33,016 instances and 169,687 instances that are selectively extracted from the KDD99 training and testing sets, respectively. The principle for forming such reduced sets is to get all instances in each small group (attack type), but only a limited amount of instances in each large group, from both the KDD99 training and testing sets. More explicitly, for forming the reduced training, we randomly select five percent of each large group *Neptune, smurf,* and *normal*, while gathering all instances in the remaining groups from the KDD99 training set. For sampling the reduced testing set, we randomly select 50 percent of each large group *Neptune, smurf,* and *normal*, whereas collecting all instances in the remaining groups from the KDD99 testing set. Class distribution of these two reduced sets is shown in Table 2.

**Table 2.** Class distribution of the reduced training and testing sets used in experiments.

| Class | Training set | Testing set | Class | Training set | Testing set |
|---|---|---|---|---|---|
| DoS | 22,867 | 118,807 | U2R | 52 | 70 |
| Probe | 4,107 | 4,166 | Normal | 4,864 | 30,297 |
| R2L | 1,126 | 16,347 | **Total** | **33,016** | **169,687** |

### 4.2 Experiment Setup

In our experiments, the predefined set $K$ is set to $\{2, 3, \ldots, 50\}$. The convergence criterion (termination tolerance) of fuzzy *c-means* clustering is set to $10^{-6}$, whereas the fuzzy degree (exponent $\alpha$ in Formulas 1-3) is set to 3. On the other hand, continuous futures are normalized by *max_min* value ranges [6]. To handle different feature types as well as express different merit contributions of features in the Euclidian space, we calculate distances between data points by the metric proposed in Formula 4.

$$d(x_i, v_j)^2 = \sum_{q}^{m} G_q \times d_q(x_{iq}, v_{jq})^2 \tag{4}$$

where $G_q$ is information gain of feature $q$ [5], and

$$d_q(x_{iq}, v_{jq}) = \begin{cases} 1, & if\ \left(x_{iq}, v_{jq} \in \{\text{symbolic}\}\right) \wedge (x_{iq} \neq v_{jq})\ or\ \left(x_{iq}, v_{jq} \in \{\text{unknown}\}\right) \\ |x_{iq} - v_{jq}|, & if\ x_{iq}, v_{jq} \in \{\text{continuous}\} \\ \frac{|x_{iq} - v_{jq}|}{t-1}, & if\ x_{iq}, v_{jq} \in \{\text{ordinals}\};\ t = |\{\text{ordinals}\}| \\ 0, & \text{otherwise,} \end{cases}$$

The base inducers ($I$) tested in our method are the C4.5 decision tree [5] and the SVM [2] with polynomial and radial basic function kernels. The feature selection technique ($\xi$) used in this experiment is Correlation-based Feature Subset Evaluation (CfsSubsetEval) with genetic search [7]. CfsSubsetEval evaluates the merit of a feature subset by considering the individual predictive ability of each feature along with the degree of redundancy between them. Those subsets that are highly correlated with the class while having low intercorrelation are preferred.

Candidate classifiers are evaluated by an attack type-based stratified cross validation ($q$=10 folds). The maximum performance classifier is determined based on overall accuracy (i.e., the ratio of the number of correctly classified instances to the total number of instances in the training set).

## 4.3 Experiment Results

The experimental comparison of our method to other well-known methods is featured in Table 3. All the compared classifiers are built from the same training set and tested on the same testing set as described in Section 4.1. Moreover, Figure 4 depicts True Positive Rates (TPRs) and False Positive Rates (FPRs) of classifiers with respect to each class label, whereas Figure 3 portrays average TPRs and FPRs of classifiers. TPR of a class $\omega_c$ is the ratio of "the number of correctly classified instances in the class $\omega_c$" to "the total number of instances in the class $\omega_c$". FPR of a class $\omega_c$ is the ratio of "the number of instances that do not belong to the class $\omega_c$ but are classified as $\omega_c$" to "the total number of instances that do not belong to the class $\omega_c$".

To have a wider comparative view, we run our algorithm (*CFC*) with different settings of two parameters (i.e., *I*: base inducer; *T*: the way to employ cluster features for building classifiers). The results of such runs are listed in Rows 10-18 of Table 3.

As shown in Figures 3 and 4, our method, in general, considerably outperforms the others with respect to TPRs in all five classes and on average. Particularly, *CFC* classifiers are significantly better than all the others in detecting hard classes (i.e., *R2L* and *U2R*). On the other hand, FPRs of *CFC* classifiers are generally lower than those of the others. Our method also considerably improves the classification ability of base inducers (SVM and DT) in both viewpoints, i.e., applying or not applying feature selection. More concretely, by using the same feature selection technique, the SVM classifier built from the manipulated training set (i.e., *CFC*(*I*=SVM,*T*=3)) is considerably superior to the SVM classifier built from the original training set (i.e., SVM_FS). Similarly, the performance of *CFC*(*I*=DT,*T*=3) is considerably better than that DT_FS. This tells that applying a feature selection technique on the manipulated training set produces a higher qualitative feature subset (including base features and cluster features) than that on the original training set.

Regarding the SVM context, although we further test PSVM (Polynomial SVM) with exponent degrees ranging from 2 to 6, its performance remains worse than *CFC*(PSVM(*degree*=2),*T*={1,2,3}). On average, *CFC*(PSVM (*degree*=2),*T*={1,2,3}) gives a 91.96% TPR (with a 2.2% FPR), whereas PSVM(*degree*={2,…,6}) produces an 86.84% TPR (with a 3.44% FPR). We also test RSVM (Radial Basic Function SVM) with widths *Gamma* ranging from 0.1 to 1.0, but its performance still underperforms *CFC*(RSVM(*Gamma*=0.1),*T*={1,2,3}). More precisely, on average, RSVM(*Gamma*={0.1,0.2,...,1}) produces an 86.72% TPR (with a 3.62% FPR), whereas *CFC*(RSVM(*Gamma*=0.1),*T*={1,2,3}) gives a 91.15% TPR (with a 2.3% FPR). This tells that cluster features benefit SVM in high dimensionality.

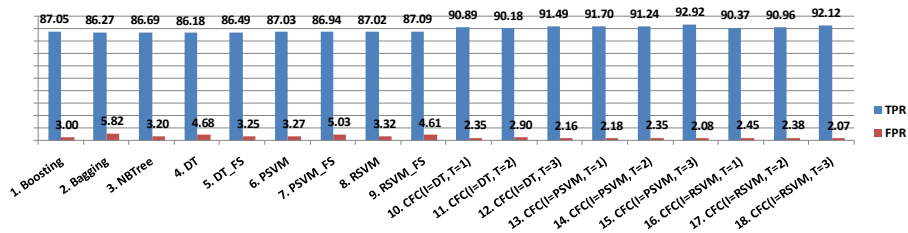

**Fig. 3.** Average True Positive and False Positive Rates (%) of classifiers

**Table 3.** True Possitive and False Possitive rates (%) of classifiers.

| Classifier | | | DoS | Probe | R2L | U2R | Normal | Average |
|---|---|---|---|---|---|---|---|---|
| 1. Boosting | | TP | 95.36 | 82.48 | 5.51 | 35.71 | 99.22 | 87.05 |
| | | FP | 0.44 | 0.46 | 0.03 | 0.01 | 15.01 | 3.00 |
| 2. Bagging | | TP | 94.72 | 80.03 | 3.46 | 42.86 | 98.74 | 86.27 |
| | | FP | 4.64 | 0.41 | 0.30 | 0.02 | 14.18 | 5.82 |
| 3. NBTree | | TP | 94.53 | 83.32 | 9.54 | 51.43 | 98.08 | 86.69 |
| | | FP | 0.84 | 0.62 | 0.60 | 0.24 | 14.22 | 3.20 |
| 4. DT | | TP | 94.72 | 78.68 | 2.84 | 51.43 | 98.77 | 86.18 |
| | | FP | 2.85 | 0.57 | 0.03 | 0.10 | 14.96 | 4.68 |
| 5. DT_FS | | TP | 94.26 | 85.60 | 7.11 | 38.57 | 99.06 | 86.49 |
| | | FP | 0.82 | 0.93 | 0.26 | 0.05 | 14.70 | 3.25 |
| 6. PSVM | | TP | 95.14 | 84.09 | 9.43 | 38.57 | 97.61 | 87.03 |
| | | FP | 0.91 | 0.52 | 0.24 | 0.01 | 14.56 | 3.27 |
| 7. PSVM_FS | | TP | 95.44 | 73.84 | 8.86 | 44.29 | 97.65 | 86.94 |
| | | FP | 3.57 | 0.22 | 0.29 | 0.01 | 14.00 | 5.03 |
| 8. RSVM | | TP | 95.11 | 83.99 | 9.51 | 38.57 | 97.62 | 87.02 |
| | | FP | 0.98 | 0.53 | 0.23 | 0.01 | 14.55 | 3.32 |
| 9. RSVM_FS | | TP | 94.98 | 81.73 | 9.92 | 40.00 | 98.64 | 87.09 |
| | | FP | 3.04 | 0.55 | 0.17 | 0.01 | 13.75 | 4.61 |
| 10. *CFC*(*I*=DT, *T*=1) | | TP | 97.69 | 88.65 | 25.93 | 58.57 | 99.62 | 90.89 |
| | | FP | 0.76 | 0.53 | 0.02 | 0.03 | 10.12 | 2.35 |
| 11. *CFC*(*I*=DT, *T*=2) | | TP | 97.46 | 88.24 | 21.47 | 60.00 | 99.03 | 90.18 |
| | | FP | 1.45 | 0.75 | 0.03 | 0.04 | 10.46 | 2.90 |
| 12. *CFC*(*I*=DT, *T*=3) | | TP | 98.30 | 90.13 | 28.01 | 62.86 | 99.27 | 91.49 |
| | | FP | 0.70 | 0.65 | 0.03 | 0.04 | 9.24 | 2.16 |
| 13. *CFC*(*I*=PSVM, *T*=1) | | TP | 98.42 | 92.49 | 28.19 | 68.57 | 99.57 | 91.70 |
| | | FP | 0.81 | 0.68 | 0.03 | 0.03 | 8.93 | 2.18 |
| 14. *CFC*(*I*=PSVM, *T*=2) | | TP | 98.12 | 92.20 | 26.27 | 75.71 | 99.20 | 91.24 |
| | | FP | 0.87 | 0.48 | 0.02 | 0.04 | 9.70 | 2.35 |
| 15. *CFC*(*I*=PSVM, *T*=3) | | TP | 98.83 | 94.89 | 37.62 | 74.29 | 99.36 | 92.92 |
| | | FP | 1.08 | 0.71 | 0.03 | 0.03 | 7.31 | 2.08 |
| 16. *CFC*(*I*=RSVM, *T*=1) | | TP | 97.42 | 95.06 | 21.61 | 68.57 | 99.23 | 90.37 |
| | | FP | 0.76 | 0.61 | 0.04 | 0.02 | 10.65 | 2.45 |
| 17. *CFC*(*I*=RSVM, *T*=2) | | TP | 98.15 | 91.72 | 22.51 | 72.86 | 99.62 | 90.96 |
| | | FP | 0.83 | 0.58 | 0.03 | 0.03 | 9.96 | 2.38 |
| 18. *CFC*(*I*=RSVM, *T*=3) | | TP | 98.22 | 94.36 | 33.81 | 68.57 | 99.45 | 92.12 |
| | | FP | 0.79 | 0.70 | 0.02 | 0.03 | 8.39 | 2.07 |

- DT refers to the C4.5 decision tree inducer [5] with established input parameters: pruning method = *pessimistic pruning*, confidence=0.2, and Min(#instances per leaf)=6.
- Boosting uses the *AdaBoost* [8] with parameters: base inducer=DT, # classifiers=10.
- Bagging uses the *Bagging* [4] with parameters: base inducer=DT, # classifiers=10.
- PSVM refers to SVM inducer with Polynomial Kernel (*exponent degree* = 2).
- RSVM refers to SVM inducer with Radial Basic Function Kernel (*width gamma* = 0.1).
- Classifiers 1-9 are trained on the original training set (without cluster features), where classifiers 5, 7, and 9 employ the feature selection technique (ξ) as described in Section 5.2, whereas classifiers 1-4, 6, and 8 do not apply the feature selection technique (ξ).
- Classifiers 10-18 are built from the *CFC* algorithm whose base inducers have the same parameter settings as *stand-alone* classifiers 4, 6, and 8.
- The column *Average* is the average weighted by the number of instances on each class.

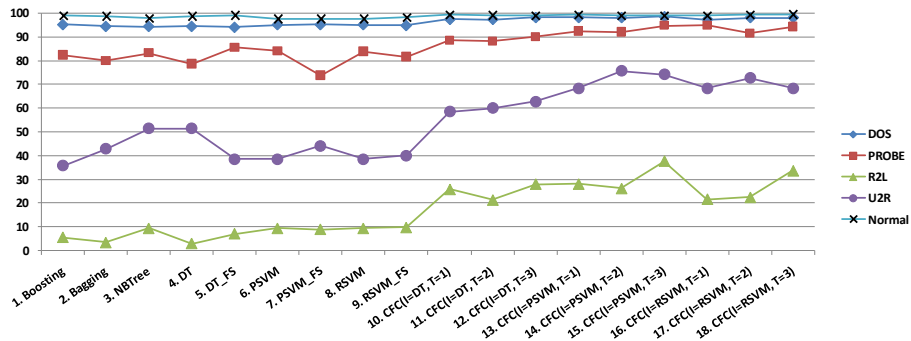

**Fig. 4.** True Positive Rates (%) of classifiers on each class.

## 5 Conclusion and Future Work

We propose in this paper a novel method in applying data mining to the intrusion detection problem. The incorporation of cluster features resulting from a fuzzy clustering into the training process is proven to be efficient for enhancing the strength of a base classifier. The tactic to achieve a high performance classifier from a training set supplemented with cluster features is addressed. We experimentally show that, as a whole, our method clearly outperforms all the tested methods. Although the experiments are conducted on the KDD99 IDS dataset, the approach we propose can be generally used to improve classification in other application domains. However, to be more objective in evaluating any data mining solution, our future work will be to test the proposed method on other real datasets. In particular, our current effort is fulfilling a honeypot system for gathering both real intrusion and normal traffic activities. Such a real dataset will then be used to evaluate the method we proposed.